\documentclass[preprint,12pt]{elsarticle}

\graphicspath{ {./figures/} }
\usepackage{subfig}
\usepackage{xcolor}
\usepackage{amssymb}
\usepackage{amsmath}

\begin{document}

\begin{frontmatter}



\title{Transition to Coarse-Grained Order in Coupled Logistic
Maps: Effect of Delay and Asymmetry}


\author[1]{Bhakti Parag Rajvaidya}
\ead{bhakti.patankar@raisoni.net}
\author[2]{Ankosh D. Deshmukh}  
\ead{ankoshdeshmukh50@gmail.com}
\author[2]{Prashant M. Gade\corref{cor1}}
\ead{prashant.m.gade@gmail.com}
\author[3]{Girish G. Sahasrabudhe} 
\ead{irigleen@gmail.com}
\cortext[cor1]{Corresponding Author}
\address [1]{G. H. Raisoni College of Engineering, Nagpur, India.}
\address [2]{Department of Physics, Rashtrasant Tukadoji Maharaj Nagpur University, Nagpur- 440033, India.}
\address [3]{Retired Professor of Physics, Shri. Ramdeobaba College of 
Engineering and Management, Katol Road, Nagpur, India.}

\begin{abstract}
We study one-dimensional coupled logistic maps with delayed linear or nonlinear nearest-neighbor coupling. Taking the nonzero fixed point of the map $x^*$ as reference, we coarse-grain the system by identifying values above $x^*$ with the spin-up state and values below $x^*$ with the spin-down state. We define persistent sites at time $T$ as the sites which did not change their spin state even once for all even times till time $T$.  A clear transition from asymptotic zero persistence to non-zero persistence is seen in the parameter space. The transition is accompanied by the emergence of antiferromagnetic, or ferromagnetic order in space.  We observe antiferromagnetic order for nonlinear coupling and even delay, or linear coupling and odd delay.  We observe ferromagnetic order for linear coupling and even delay, or nonlinear coupling and odd delay. For symmetric coupling, we observe a power-law decay of persistence. The persistence exponent is close to $0.375$ for the transition to antiferromagnetic order and close to $0.285$ for ferromagnetic order. The number of domain walls decays with an exponent close to $0.5$ in all cases as expected.  The persistence decays as a stretched exponential and not a power-law at the critical point, in the presence of asymmetry.
  \end{abstract}




\end{frontmatter}


\section{Introduction}
\label{}
Phase transitions have been an intriguing and important class of physical phenomena
to many workers for several years. These phenomena involve a drastic transformation
in macroscopic properties of matter as certain parameters change. The system is said to
change phase. The system undergoing a phase transition may have a critical 
manifold separating region in an appropriate parameter space such that for parameter
values in any one region the system is in one phase. For example, a liquid-gas
transformation is described by a phase diagram delineating such regions in the p-T
plane. In this case, two phases – e.g., liquid and gas – coexist on the separating line.
The coexistence of phases may not occur 
in all instances of phase transitions \cite{jaeger1998ehrenfest}. Transitions
with the coexistence of phases belong to a class called first order phase transitions. Certain
magnetic systems show a second order phase transition from the paramagnetic to the
ferromagnetic state at a critical temperature. In this case, the second order derivatives of
free energy undergo a discontinuous change. We do not observe coexistence in this case.
As is well known, certain power laws describe the behavior of these second derivatives as
the transition temperature is approached. The exponents, called critical exponents, in
these power laws, are 
universal \cite{martin2001} and describe the magnetic transition for all materials
which are capable of it. 

Although the magnetic transition is an equilibrium transition at the critical temperature,
similar phenomena are observed in non-equilibrium systems too. Coupled map lattices
(CML) \cite{kaneko1984period} constitute a 
very useful class of models to study non-equilibrium systems. These models
are computationally cost-effective and more well-organized compared 
to other such
models. The models can perform time evolution of systems with both  
linear or
non-linear interactions. CML models are discrete-space, discrete-time models \cite{kaneko1984period}, in which
real values or vectors are assigned to each point in a lattice of 
points in space. The time evolution of the system is 
defined by dynamical equations of the model, which are in the
form of update rule for site vectors.

The original CML model put forward by Kaneko was defined by \cite{kaneko1984period},
\begin{equation}
	x_{i}(t+1)=(1-\epsilon) f(x_{i}(t)) + {\frac{\epsilon}{2}}[f(x_{i-1}(t)) + f(x_{i+1}(t))]
\end{equation}

It incorporates the nearest neighbor (NN) interactions with coupling constant ε. Here $x_{i}(t)$ is
the real value attached to the $i^{th}$ lattice at time t. The function f(x) is generally a non-linear
function of x. This basic model has been extended in various subsequent studies to include
feedback, global interactions, and stochastically determined couplings. Transitions in such
CML systems address changes in the global state of the lattice, often in asymptotic 
time (i.e., as $t\to\infty$). 

This system without delay has been
studied extensively. The original motivation for these studies has been ‘field theory of
turbulence’ and the argument was that we can build our understanding of high
dimensional spatiotemporal chaos drawing from the knowledge of low dimensional
chaotic systems. The studies have been mainly numerical and several patterns have
been obtained in this simple system. The spatiotemporal patterns are visually 
identified
and several phases have been classified in this system. 
There have been relatively fewer studies on 
the
effect of delay\cite{marti2003delay,masoller2005random,atay2004delays}.

In this work, we have coarse-grained the variables and studied persistence in
the coupled map lattices. Persistence in spatially extended systems
has been studied before \cite{derrida1994non} and can
be useful in identifying phases\cite{lemaitre1999phase,kockelkoren2000phase,tucci2003phase}. 
In particular, fully or partially
arrested states can be identified using persistence.
In general, persistence signifies that as a system evolves in time it retains some
particular property till time $t$. The property can, for example, be the sign of a variable $x$
of the system. Persistence in the context of stochastic processes is defined by examining
when a stochastically fluctuating variable crosses a threshold value for the first time.
Persistence tells us how a system retains its memory as it evolves. Often one defines a
persistence probability $P(t)$. 
If $P(t)$ follows a power-law like $P(t)\propto 1/t^{\theta}$, $\theta$ is called the persistence exponent.
Persistence exponents for stochastic systems have attracted attention \cite{jabeen2007universality} 
as a new class
of exponents exhibiting remarkable universality, like that in Ising, or Pott’s models.

Transitions to spatial intermittency and spatiotemporal intermittency have been
observed in coupled circle maps. The transition to spatiotemporal intermittency is found to
be in directed percolation (DP) universality class \cite{jabeen2007universality}. 
In 2-d coupled map lattice, Miller and
Huse showed that the transition to ‘ferromagnetic’ state is in the Ising class for a specific class of
maps \cite{miller1993macroscopic}. There has been debate on how the nature of transition changes with synchronous
or asynchronous update \cite{marcq1997universality}. Even prior to these works, Oono and Puri studied coupled
dynamical systems, which they call cell dynamical systems. They studied phase separation
dynamics using these systems as model \cite{oono1988study}. Some coupled map lattices have been claimed
to be in the universality class of Potts model \cite{salazar2005critical}. Recently, Gade and coworkers have studied
transition to ‘chimera’ type states using persistence as an
order parameter in coupled map
lattice \cite{mahajan2010transition,sonawane2011dynamic,mahajan2018stretched}.

The next section describes the extensions of Kaneko\textsc{\char13}s basic CML model, which we investigate here. 
The concept of persistence we use is introduced. Section 3 
details the results symmetric coupling. We show the bifurcation 
diagram as well as phase-space plots. We demonstrate the
existence of long-range order at the critical point
and discuss how the persistence exponent is universal for
and dependent only on ferromagnetic or antiferromagnetic order.
In the next section, we discuss asymmetric coupling.
In the final section, we give discuss the results and underline our
major findings.

\section{The Model}
The CML models studied here are logistic map CMLs, where the non-linear function
is the logistic map $f(x)= \mu x(1-x)$. The parameter $\mu$ ranges over the chaotic region of the logistic map.
We introduce linear, or non-linear delayed NN-coupling into the lattice update
equation. The general model is defined by,

\begin{equation}
	x_{i}(t+1)=(1-\epsilon) f(x_{i}(t)) + {\frac{\epsilon}{2}}[x_{i-1}(t-\tau ) + x_{i+1}(t-\tau )] 
\end{equation}

for linear NN Coupling, and by 
\begin{equation}
	x_{i}(t+1)=(1-\epsilon) f(x_{i}(t)) + {\frac{\epsilon}{2}}[f(x_{i-1}(t-\tau )) + f(x_{i+1}(t-\tau ))] 
\end{equation}
for the non-linear case.

Thus, the non-linearity in coupling is defined by the same function f. 
Here, $\epsilon $ is the NN-coupling strength, and $\tau $, the delay, or time lag in the
NN coupling. The index $i$ ranges over 1 to N, the total number of lattice sites, or the
lattice size. As is usual, we impose cyclic boundary conditions, where the last lattice
point is the neighbor of the first. The variables $x_{i}(t)\in [0,1] $ are the real values attached 
at time-step $t$
to the $i^{th}$ lattice point of a one-dimensional lattice of size $N$. The 
nonzero fixed point of the
logistic map is given by,

\begin{equation}
x^{*}=1-1/\mu
\end{equation}

Simple analysis using $x^*=1-\frac{1}{\mu}$
shows that $f(x)-x^*=(x-x^*)(2-\mu(1+x-x^*)) =
		     -\mu(x-x^*)((1-2/\mu)+(x-x^*)) $.
It
follows that $x$ and $f(x)$ are always on opposite side of $x^*$ provided
$(x-x^*)>0$ or 
$\vert x-x^*\vert <1-\frac{2}{\mu}$. (We assume $\mu>2$.)

The system with asymmetry was introduced by Kaneko\cite{kaneko1985spatial}
We further modify this model to introduce the partial symmetry breaking coupling and/or
nonlinearity. The modified equation is as below: 

\begin{equation}
	x_{i}(t+1)=(1-\epsilon) f(x_{i}(t)) + (\frac{\epsilon}{2}+D)[x_{i-1}(t-\tau)]+ (\frac{\epsilon}{2}-D)[x_{i+1}(t-\tau)] 
\end{equation}
for linear case while

\begin{equation}
	x_{i}(t+1)=(1-\epsilon) f(x_{i}(t)) +(\frac{\epsilon}{2}+D)
	[f(x_{i-1}(t-\tau))]+ (\frac{\epsilon}{2}-D)[f(x_{i+1}(t-\tau))]
\end{equation}
for non-linear case.

Where, $0< D < {\frac{\epsilon}{2}}$. We assume periodic boundary conditions. 
For $D\ne 0$, the coupling is aymmetric.
The value of D is taken as $5\%$ of $\epsilon$.

\section{Symmetric Coupling: Bifurcation Diagram and Long-Range Spatial Ordering }

We simulate the above system keeping $\mu=3.9$ and varying $\epsilon$
and delay ranging from 0 to 4. 
(Similar results are obtained for other values of $\mu$.)
This is symmetric coupling and $D=0$.  For each value of
$\epsilon$ and fixed delay,  we start with randomized lattice values
for $N=100$
and evolve them for large time $t>8 \times 10^5$. We plot
these asymptotic lattice values at all sites against $\epsilon$. Fig. 1
shows the result.
For very small coupling, we observe that the lattice values are
confined to a single band. When the coupling $\epsilon$ is greater
than critical coupling $\epsilon_c$, we observe that the  lattice
values are confined to two different bands. The 
value at a given site alternately visits the two bands.

\begin{figure}[ht!]
	\centering\includegraphics[width=5.5in,height=5.5in]{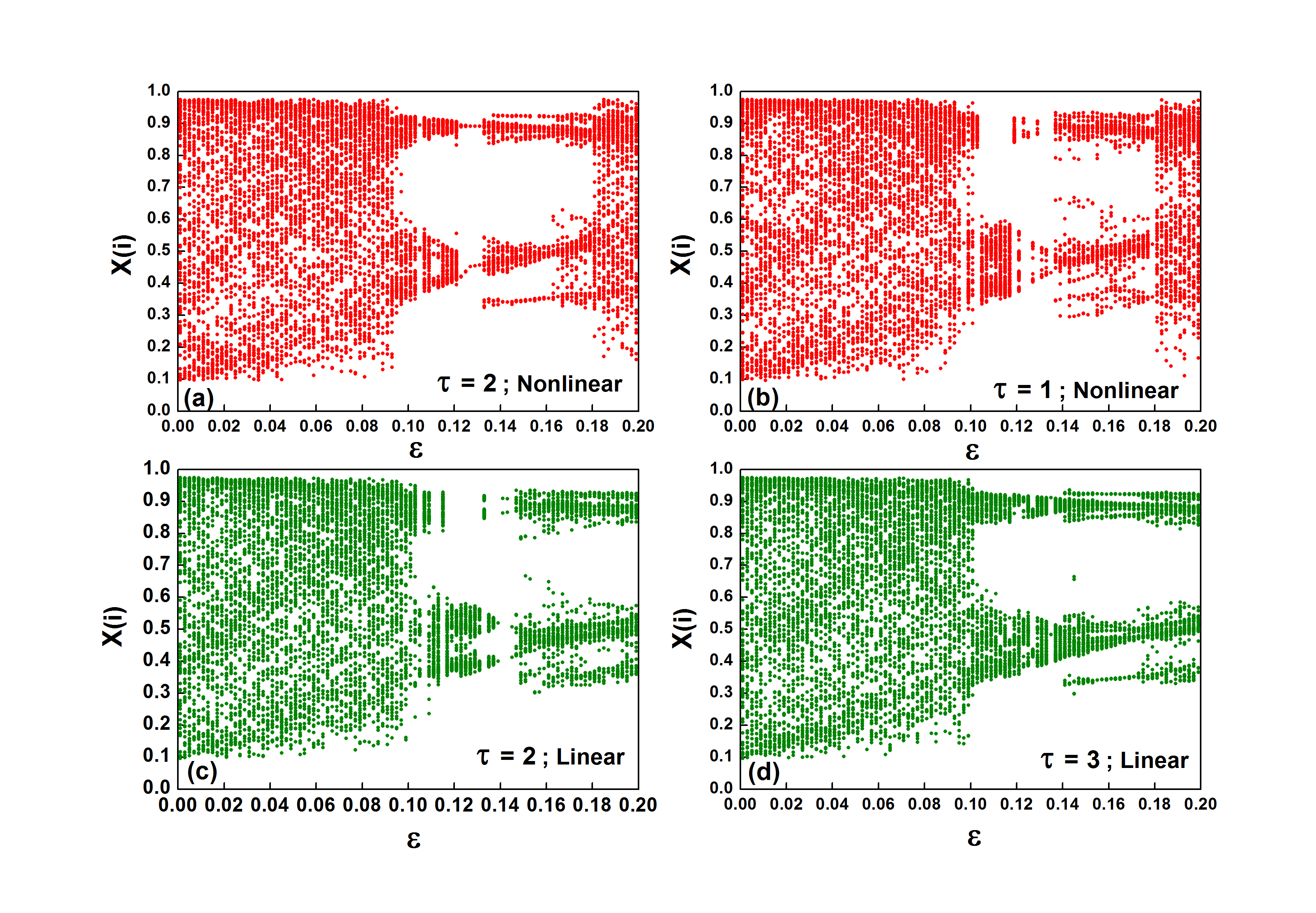}
	\caption{The bifurcation plot is plotted as a function of 
	$x(i)$ versus $\epsilon$ for nonlinear system with 
	delay: (a) $\tau = 2$ 
	and (b) $\tau =1$, and for linear system with delay: (c) $\tau = 2$ 
	and (d) $\tau =3$. We have shown only representative cases. Similar
	bifurcation diagrams are obtained for all values of $\tau$.}
	\label{fig1}
\end{figure}

Now, we decide to assign spin value $s_i(t)=1$ (spin-up state) to variable $x_i(t)$ 
if $x_i(t) > x^*$, and the spin value $s_i(t)=-1$ (spin-down state) 
if $x_i(t) < x^*$. 
In other words, the two bands in Fig. 1 correspond to distinct spin states. 
Initial conditions are random and all sites are in either of the two spin 
states (that is, the site value are $< x^*$, or $\ge x^*$) to begin with. When the system is 
asymptotically stuck in a two-band state (as in a part of $\epsilon > \epsilon_c$ 
region in Fig. 1), we also find that a $\emph{finite fraction}$ of initial lattice 
site values have all the time been stuck in their $\lq$starting-band'
at even times.
The spin 
values at these sites have strictly 
alternated between $\pm$ during the entire evolution till then. 
This is called persistence.
The values in each part are expected to return to the same part after two
applications of the logistic map. We now extend the same 
expectation to the application
of the dynamical equations of Eq.(1) to the CML values.  
We start with a randomized set of lattice values $x_{i}(0) $ 
and random past lattice values for the nonzero delay. 
(The quantitative results on lower critical line do not change 
much for zero past values.)
We examine the spin values at all even time-steps.
We say that the site $i$ is persistent till time $t$ if 
$s_i(2t^{'})=s_i(0)$ for all times $t^{'}$ such 
that $0\le t^{'} \le t $. 
In other words, the sites which did not change their spin value 
{\em{even once}} at all even time-steps are persistent sites.
The fraction of such persistent sites, denoted by $P(t)$, is called 
persistence at time $t$. This definition of
persistence derives from local spin persistence.
There is a significant part of phase space where the persistence is nonzero.
We have shown the values of $\epsilon$ and $\mu$ for which asymptotic value
of persistence is nonzero in Fig. 2.
We have shown the phase-plot for $\tau=1,2$ for nonlinear coupling and for $\tau=2,3$
for linear coupling. Similar figures
are obtained for other values of $\tau$.
In this work, we specifically consider the transition to nonzero persistence
for $\mu=3.9$ on the lower critical line.

\begin{figure}[ht!]
	\centering\includegraphics[width=5.5in,height=5.5in]{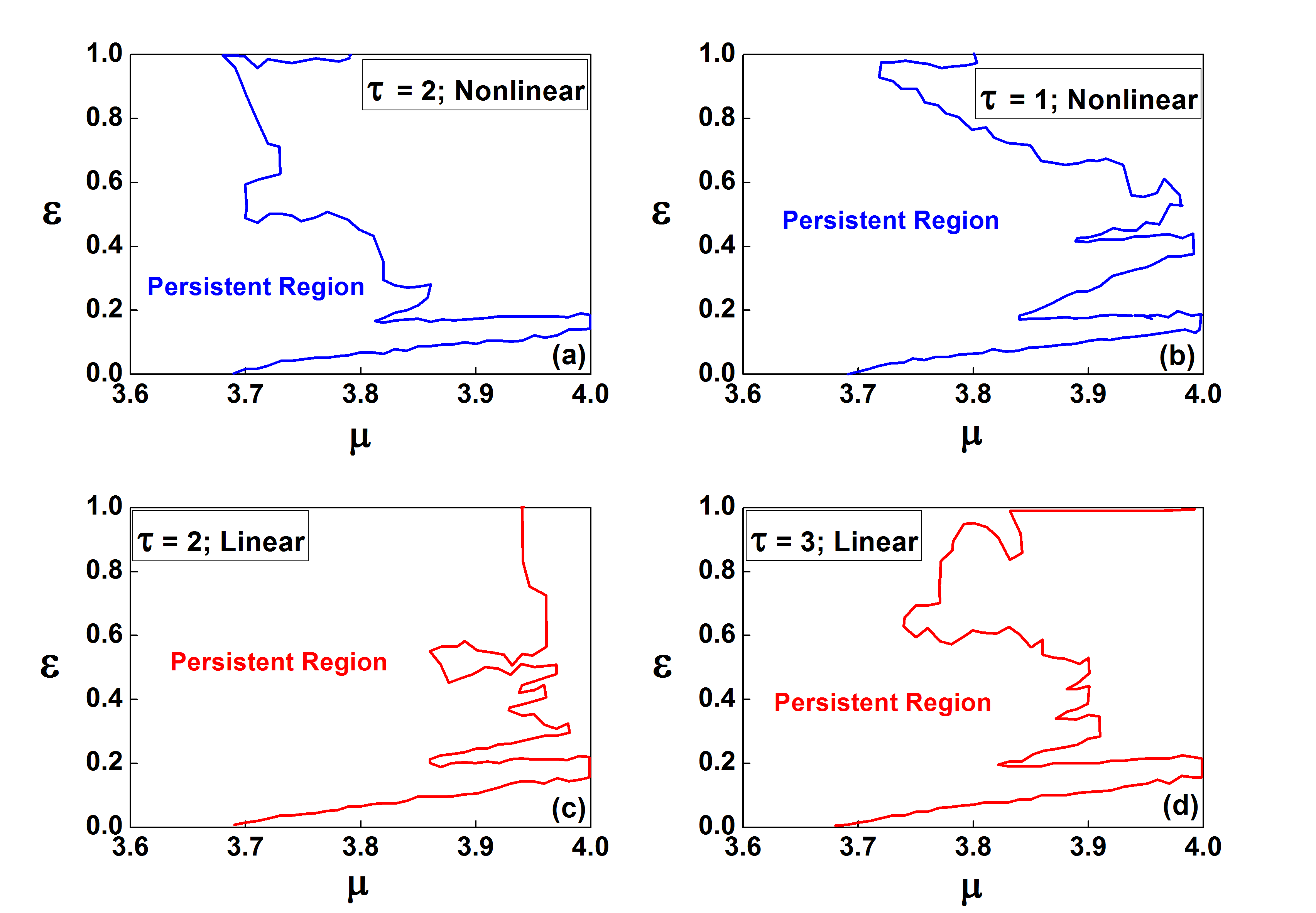}
	\caption{The interface of persistent regions are
	plotted for nonlinear symmetric coupling for
	delay: (a) $\tau = 2$, and (b) $\tau = 1$ 
	and for linear symmetric coupling for
	delay: (c) $\tau = 2$, and (d) $\tau = 3$. 
	The approximation by compact
	region is schematic and 
	there are some persistent points  outside the persistent 
	region.}
	\label{fig2}
\end{figure}

In magnetic systems, the motion of domain walls leads to a
ferromagnetic or antiferromagnetic asymptotic state. In one dimension,
we do not expect long-range order. However, an interesting 
behavior emerges on the lower critical line.

Though the  temporal behavior of switching between two chaotic
bands is common for both types of coupling and
any delay, we observe two distinct sub-classes if we take a snapshot of
spatial profile in two-band attractor state at the critical point.

In case a), we find that odd and even sublattice are in different bands at a 
given time step
and
in  case b), all sites in the lattice are in the same band. (See Fig. 3)
In both cases, all sites move to other band at
next time step and return back to same band after two
time steps.
They keep returning to the same band at all even times.
and make a transition to the other band at all odd times.
We call a) and b) as states with antiferromagnetic and ferromagnetic
order respectively.
Using the above analogy of spin values,
we expect that $M(t)=\sum_{i=1}^N s_i(t)$ acquires
a nonzero value in ferromagnetic
state while for antiferromagnetic state $M'(t)=\vert \sum_{i=1}^{\frac{N}{2}} (s_{2i}(t)-s_{2i-1}(t))\vert$
acquires a nonzero value.
\begin{figure}[ht!]
	\centering\includegraphics[width=5.5in,height=5.5in]{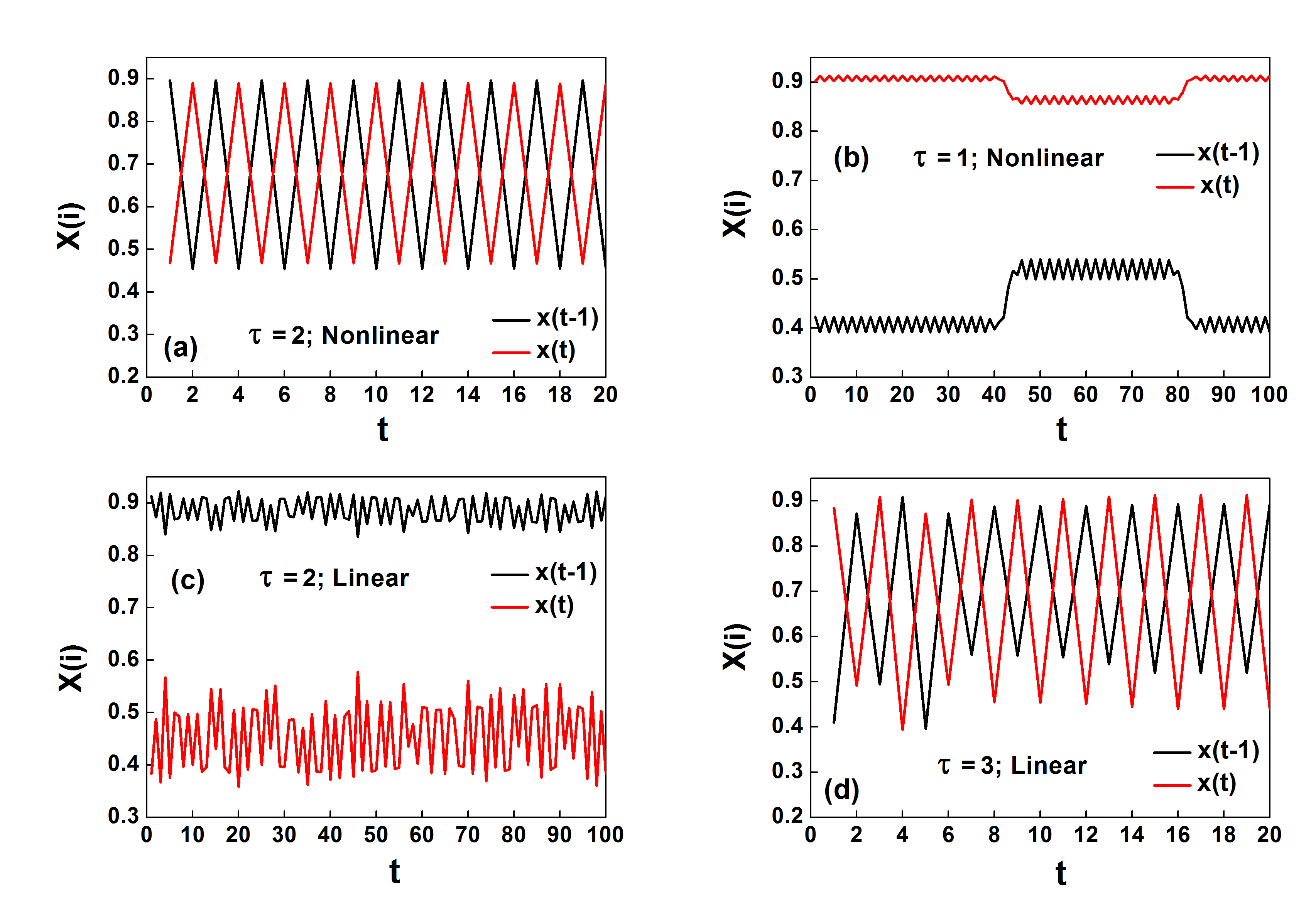}
	\caption{A spatial profile $x_i(t)$ (Red colour) and 
$x_i(t-1)$ (Black colour)
	is plotted as a function 
	of site label $i$ at large $t$ at $\epsilon=\epsilon_c$ for
	(a) nonlinear coupling and $\tau=2$ (b) nonlinear coupling and
	$\tau=1$ (c) linear coupling and $\tau=2$ and (d) linear coupling
	and $\tau=3$. We observe a zig-zag, antiferromagnetic
	pattern in (a) and (d) and all sites are confined to single
	band in (b) and (c). There are no domain walls and 
	all sites visit another band at next
	time-step.
	}
	\label{fig3}
\end{figure}

Such a state in one dimension will essentially imply the absence of
domain walls. 
A domain wall in one dimension for antiferromagnetic order is
appearance of two consecutive {\emph{like}} spins. For 
ferromagnetic order, it is appearance of two consecutive {\emph{unlike}}
spins.
In magnetic systems, the motion of domain walls leads to ferromagnetic or 
antiferromagnetic order. 
These walls are expected to undergo random walk and 
upon meeting each other mutually annihilate
thus asymptotically leading to
long-range order. 

This is precisely the picture for both ferromagnetic 
and antiferromagnetic ordering, which we observe (See Fig. 4). 
However, we observe this long-range order at a certain critical point only.
Above the critical point, domain walls stop moving and get localized.
Below the critical point, they do not freeze. But their
number still does not go to zero.
The number of domain walls saturates below as well as above the critical point.
We investigate the transition 
from
spatiotemporal chaos to 
this frozen state via a state of long-range order.

\begin{figure}[ht!]
	\centering\includegraphics[width=5.5in,height=5.5in]{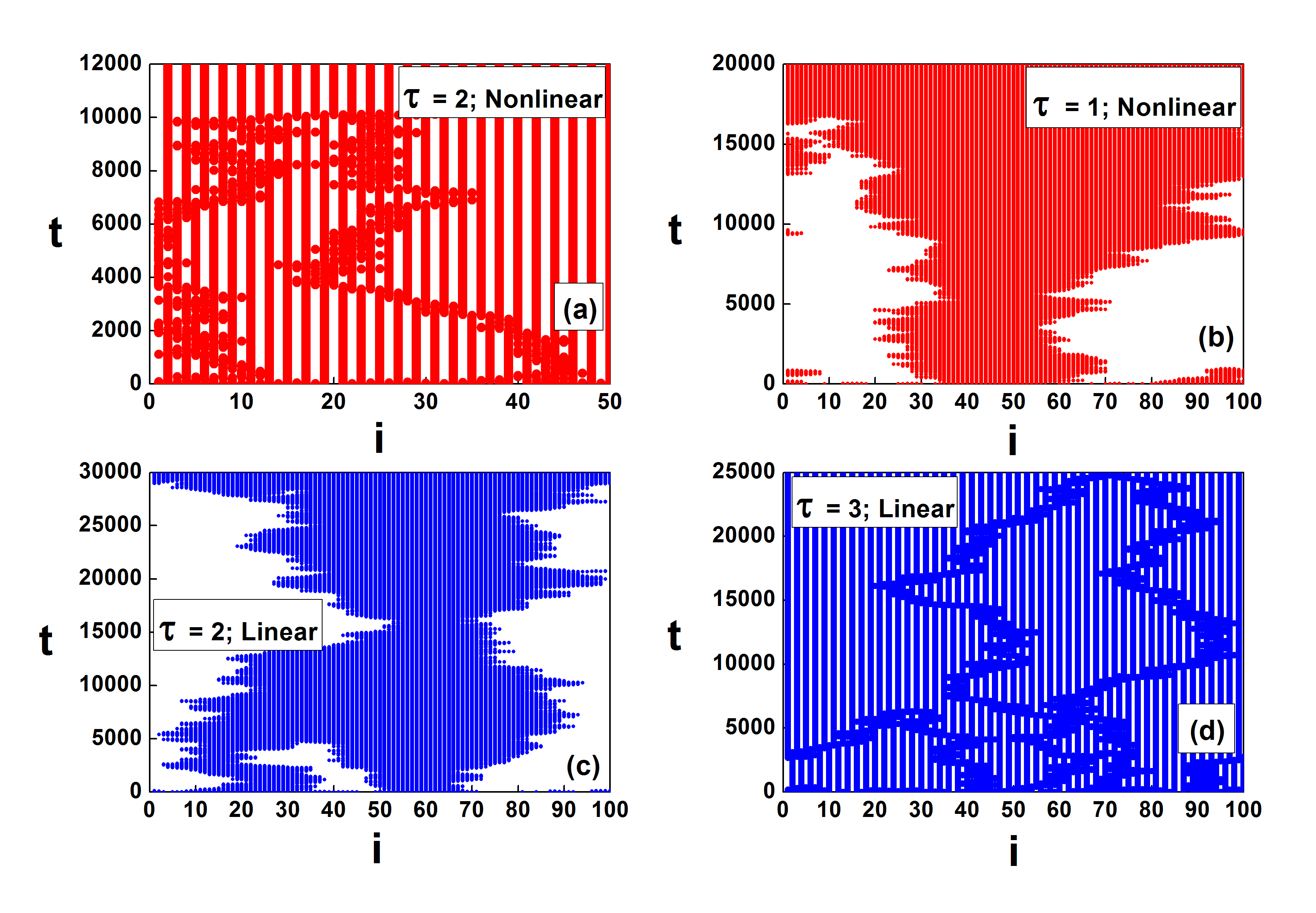}
	\caption{Domain walls motion leading to long
	range order in one dimension (a) nonlinear coupling, $\tau=2$ 
	(b) nonlinear coupling, $\tau=1$ (c) linear coupling,
	$\tau=1$ and (d) linear coupling, $\tau=3$.
	One of the spin values is marked in black.
	We observe antiferromagnetic ordering for nonlinear coupling with 
	even delay or linear coupling with odd delay. On the other hand,
	we observe ferromagnetic ordering for nonlinear coupling with
	odd delay or linear coupling with even delay.
		}
	\label{fig4}
\end{figure}

\section{Persistence and Domain walls at the critical point}

Finding an appropriate order parameter is extremely useful in studies of phase
transitions. In our case, we find a single scalar that is 
nonzero in the phase we are
interested in and zero everywhere else. This helps us 
to distinguish between different
phases without having to visually identify them. Also, a 
quantitative study of such order
parameter can give valuable information on the nature of phase transitions. 

We study two quantities, i) Density of domain walls and 
ii) Persistence to quantify the
transition. 
As mentioned above,
if $s_i(t)\ne s_{i+1}(t)$, we call
it a domain wall at site $i$ for
ferromagnetic ordering, while for antiferromagnetic ordering, if $s_i(t)
=s_{i+1}(t)$, we call it a domain wall at site $i$. We expect the order 
parameter, {\it {i.e.}}, density of domain walls to decay as a power-law with
exponent 1/2 at the critical point assuming that the comparison with
Ising model is valid. This behavior is expected if domain walls undergo
random walk and merge upon meeting. This quantity goes to zero at 
the critical point as a power-law but
has a nonzero steady-state value above or below the critical point.

The other quantity 
persistence, $P(t)$ is the fraction of
sites, for each of which the spin at all even times till time
$t$ is the same as the initial spin. 
It has been found useful for 
studying the transition to fully or partially arrested states in
crisis in coupled map lattices.

The non-linear case in Eq.(3) with $\tau = 0$, i.e., without delay has
been extensively studied by two of us \cite{gade2013universal}, 
and preliminary investigations on the case with delay are presented in
\cite{rajvaidya2019}. 
We extend the
study to cases of different delay values as well as those with linear and nonlinear coupling. 

Let $\epsilon_c$ be the value of coupling at a critical point. We 
define a critical point $\epsilon_c$ as a point such that for $\epsilon<\epsilon_c$,
the value of $P(t)$ is zero asymptotically. We focus on a critical line of critical points 
closest to the $\mu$-axis in the phase plots.
We plot
the behavior of $P(t)$ as a function of $t$ for  i) $\epsilon<\epsilon_c$ ii) 
$\epsilon=\epsilon_c$ and iii) $\epsilon>\epsilon_c$.
The persistence goes to zero quickly below the critical point and
saturates above the critical point. On the other hand, the density of domain 
walls $d(t)$ saturates both above and below critical point and shows
power-law with exponent close to half only at the critical point. 
The order parameter can distinguish a phase if it is positive only
within the phase and zero outside. In this sense, persistence clearly
acts as an order parameter for the crisis state. 

\begin{figure}[ht!]
	\centering\includegraphics[width=5.5in,height=3.5in]{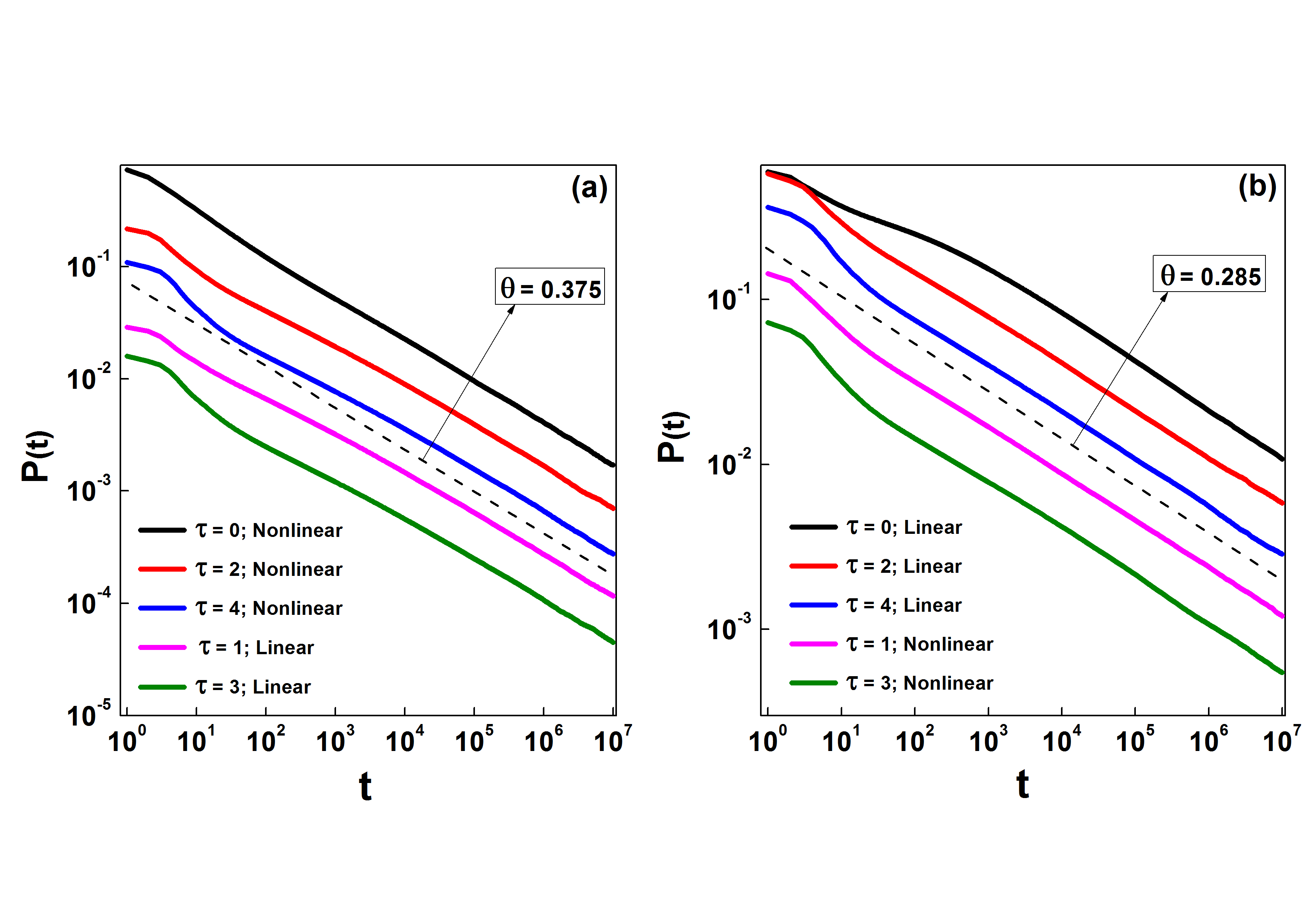}
	\caption{The persistence $P(t)$ is plotted as a function of 
	time $t$ for (a) nonlinear coupling with delay values 
$\tau =0, 2, 4$ and linear 
	coupling with delay values $\tau =1, 3 $ (from
top to bottom)
	and for (b) linear coupling with delay values $\tau =0, 
2, 4$ and nonlinear coupling with delay values 
$\tau =1, 3$ (from top to bottom). 
The Y-axis values ($P(t)$) are multiplied by arbitrary constant for better 
visualization. For all cases, $N=10^5$. We average over 20 configurations.
	}
	\label{fig5}
\end{figure}
\begin{figure}[ht!]
	\centering\includegraphics[width=5.5in,height=3.5in]{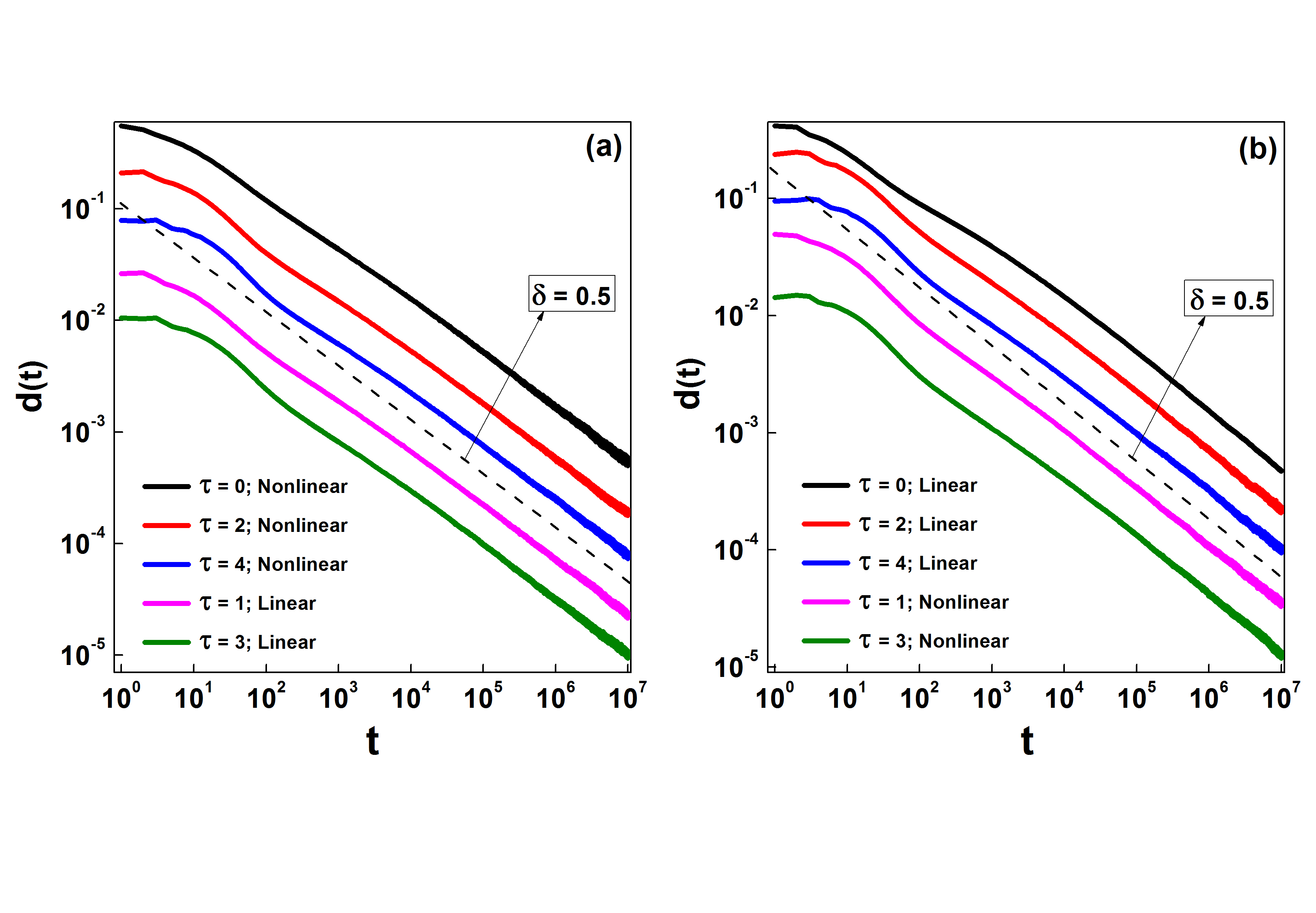}
	\caption{The domain walls $d(t)$ is plotted as a function of time 
$t$ for (a) nonlinear coupling with delay values $\tau =0, 2, 
4$ and linear coupling with delay values $\tau =1, 3$ (from top
to bottom) and for (b) linear coupling with delay values $\tau =0, 2, 
4$ and nonlinear coupling with delay values $\tau =1, 3
$ (from top to bottom). 
The Y-axis values ($d(t)$) are multiplied by arbitrary constant for better 
visualization. For all cases, $N=10^5$. We average over 20 configurations.
	}
	\label{fig6}
\end{figure}

For nonlinear case, we find that 
i) 
$\epsilon_c= 0.11$, for $\tau=0$, 
ii) 
$\epsilon_c= 0.105$, for $\tau=1$, 
iii) 
$\epsilon_c= 0.11$, for $\tau=2$, 
iv) 
$\epsilon_c= 0.11$, for $\tau=3$, 
and v) 
$\epsilon_c= 0.11$, for $\tau=4$. 
For linear case, we find that i) 
$\epsilon_c= 0.115$, for $\tau=0$, 
ii) 
$\epsilon_c= 0.11$, for $\tau=1$, 
iii) 
$\epsilon_c= 0.115$, for $\tau=2$, 
iv) 
$\epsilon_c= 0.11$, for $\tau=3$, 
and v) 
$\epsilon_c= 0.115$, for $\tau=4$. 
We study the dependence of persistence exponent on delay time and nonlinearity. We
observe that the persistence
exponent is dependent only on a combination of two factors: a) Nature of 
coupling: linear, or nonlinear 
b) delay: odd, or even.
For nonlinear coupling and zero or even time-lag, we
observe antiferromagnetic ordering. For linear coupling, and odd
time-lag, we observe antiferromagnetic ordering. The
persistence exponent is ${\frac{3}{8}}=0.375$ for all these 
cases (See Fig. 5(a)).
This is a persistence exponent for Ising model\cite{derrida}. 
On the other hand for linear coupling and zero or even time-lag or 
nonlinear coupling and odd time-lag, we observe
ferromagnetic ordering. The persistence exponent
is $0.285\sim {\frac{2}{7}}$ in all these cases (See Fig. 5(b)).
Thus the system shows remarkable universality despite extensions 
introduced in the model.
The number of domain walls decays with an exponent close to $1/2$ in all these
cases (See Fig. 6(a) and 6(b)).

The comparison with
Ising model dynamics in one dimension is very appropriate 
since ${\frac{3}{8}}$ is Ising-type persistence exponent for
antiferromagnetic order\cite{derrida}. 
The behavior described above could be understood as follows: 
The motivation for defining persistence based on a modulo 2 
(rather than modulo 1!) dynamics was that a single application of 
logistic map flips the spin state of a typical 
lattice site and a double application typically retains the 
spin state. (The slope $f'(x^*)$ at unstable fixed point $x^*$ is negative.) 
However, (i) since there is a finite probability that a spin down state 
does not change to a spin-up state under logistic map, and (ii) since the 
dynamical equations also involve other terms like 
the coupling, clearly, retention of spin state will not 
happen for a certain non-zero fraction of sites. 
Thus, if the lattice or  a part of it is stuck in 2-band 
attractor state, a certain fraction of sites will not flip spin 
even asymptotically at each modulo 2 dynamical-step. 
If it is not stuck in 2-band state, every site will eventually flip 
leading to zero persistence. 
At the critical point or below it, we  observe the 
decay of persistence with time.

The nature of this decay is expected to depend on 
parameters of the system such as the coupling strength and 
the time lag. Our computation shows that coupling with the logistic map 
(i.e. non-linear coupling) acts as an effective negative spin-coupling for 
the lattice for long-time evolution. The dynamics, as well as qualitative 
features are then similar to the Ising model. 
Linear coupling is like coupling with the identity map and it acts as
positive coupling. Long-range ordering in this system is different
and so is persistence exponent.
Since  the coupling is small, value of $x_{i-1}(t)$ is close to 
$f(x_{i-1}(t-1))$. Thus linear 
coupling without delay is similar to
nonlinear coupling with $\tau=1$.
Thus a nonlinear system with delay $\tau=1$ acts like linear
system without delay. With even delay, the map is iterated even number
of times. 
For nonlinear coupling, with zero delay, the bands tend to alternate
since $f'(x^*)$ is negative. For $\tau=1$, the inverse image 
of map with two iterations is of interest. The sign of $f'(x^*)$ is same
as sign of $f^{-1'}(x^*)$. (Because $dy/dx={\frac{1}{(dx/dy)}}$).
Since $x^*$ is a fixed point, 
$f^{(\tau+1)'}(x^*)=(f'(x^*))^{(\tau+1)}$, and its sign is negative
for even delay and positive for odd delay.
So the system with 
even delay acts like system with no delay for nonlinear coupling.
As we argued above, system with odd delay and nonlinear coupling
is like system with even delay and linear coupling.
Thus, coupling to delayed lattice value with odd delay can
convert a positive spin-coupling to a negative one and conversely.

Therefore, in general, non-linear coupling with even delays stabilizes, and 
grows subsets of sites in which the neighboring sites are in different 
bands, i.e., have values on opposite sides of $x^*$. 
The same is true for linear coupling with odd delays. Similarly, non-linear 
coupling with odd delays, as well as linear coupling with even delays, 
stabilize and grows subsets with neighboring sites in the same band.
Though we can thus qualitatively decipher the ferro or antiferro 
ordering with these
hand-waving arguments, it is a surprise that it turns out so well for
all values of delay $\tau$ studied by us from $\tau=0$ to 4.

\section{System with asymmetry}
So far, we have studied the system at $D=0$ in Eqs. 5 and 6. 
The phase plots do not change much in the
presence of small asymmetry. However, the disturbances
move towards left or right and persistence decays faster. 
There is no well-defined
persistence exponent at the critical point. 
The persistence may decay as stretched
exponential or exponential. 
For example, for $\tau=1$ and $\tau=2$ and  $D=0.05\epsilon$, 
we observe stretched exponential
decay of persistence at critical point for 
nonlinear coupling. (See Fig. 7). 
In this case $P(t) \sim \exp(-rt^{0.5})$. However,
the behavior depends on asymmetry and delay. The behavior at the critical
point is not uniform at all critical points in the presence of asymmetry.

\begin{figure}[ht!]
	\centering\includegraphics[width=3.5in,height=3.5in]{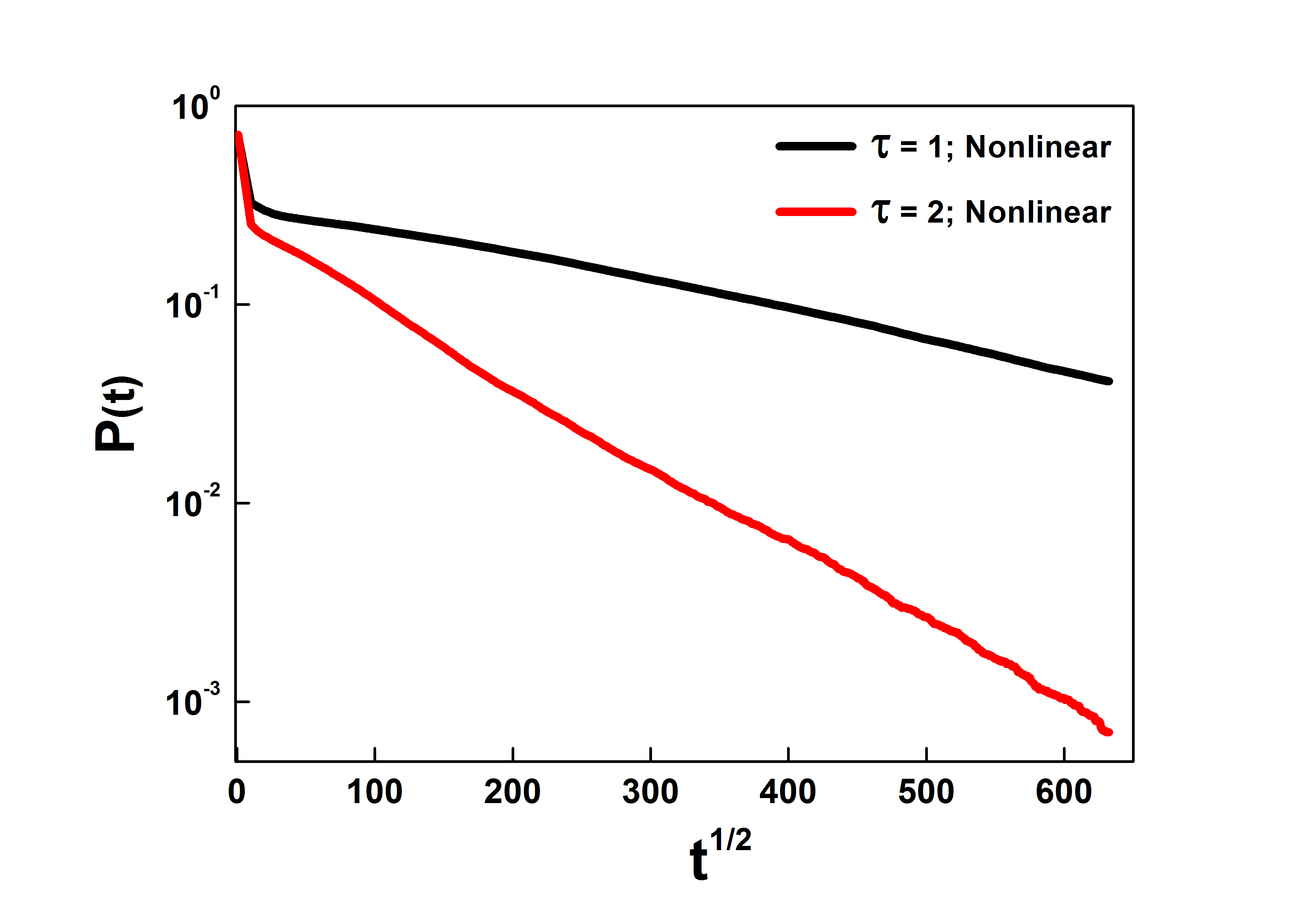}
	\caption{The stretched exponential decay is observed with  
	exponent $0.5$ for nonlinear coupling with 
	$\tau = 1$ ($\epsilon_c = 0.1423$) and $\tau = 2$ ($\epsilon_c = 0.1355$) and $D=0.05\epsilon$.
	}
	\label{fig7}
\end{figure}


\section{Discussion}
In all cases of symmetric coupling
discussed above, as the critical line is crossed, 
we obtain a transition to a
state with long-range ferromagnetic, or antiferromagnetic order. 
We conclude that
linear coupling with zero or even delay acts like ferromagnetic 
coupling, whereas, nonlinear coupling with zero, or even delay acts 
like antiferromagnetic coupling. On the
other hand, linear coupling with odd delay leads to antiferromagnetic order, and
nonlinear coupling with odd delay leads to ferromagnetic order. Also, on the critical line
time-decay of persistence is governed by a power-law. 
For antiferromagnetic order,
persistence exponent is $\approx 3/8$ while for ferromagnetic order the persistence exponent is $\approx 2/7$.

These results are surprising for the two reasons a)
Persistence exponents are known to change with detailed dynamics of the system. 
But the exponents take only two values depending on the asymptotic
state. b) The exponent $0.375$ is obtained for antiferromagnetic order. This exponent $0.375$
is reported for the Ising model. 
It is interesting that, for all cases with antiferromagnetic
order, the same exponent is obtained. This underlines 
the similarity of coupled logistic maps
with nonlinear coupling with Ising model. Surprisingly, the exponent changes 
for
the system leading to ferromagnetic order. For the 1-d Ising model, 
the system with negative
coupling $J$ can be mapped to the system with positive coupling by changing 
the sign of spins of
one of the sub-lattices. Thus there is no real difference in these two 
cases for the Ising
model. However, we obtain Ising-like exponents only when there is antiferromagnetic
order asymptotically. The exponent in the other case is also universal. Nevertheless, it is not
reported before.

In presence of asymmetry, there is no universal behavior at the critical
point. The decay is faster than power-law. However, the detailed behavior
depends on specific parameters such as the strength of asymmetry.


\section*{Acknowledgments}
ADD and PMG thanks DST-SERB for financial assistance (EMR/2016/006685). 


 \bibliographystyle{elsarticle-num} 
 \bibliography{new.bib}




\end{document}